\documentclass{iau}
\usepackage{graphicx} 
\usepackage{amsmath}

\title[IAUS291.~~Globular cluster pulsar luminosity function] %% short title %%
{Constraining the luminosity function parameters and population size of radio
pulsars in globular clusters} %% full title %%

\author[J. Chennamangalam et al.\ ]  %% short author list %%
{Jayanth Chennamangalam$^1$,
D. R. Lorimer$^{1,2}$,
Ilya Mandel$^3$
\and Manjari Bagchi$^1$
}

\affiliation{$^1$Department of Physics, West Virginia University,
  \\ PO Box 6315, Morgantown, WV 26506, USA \\
email: {\tt jchennam@mix.wvu.edu}\\[\affilskip]
$^2$NRAO, Green Bank Observatory, PO Box 2, Green Bank, WV 24944, USA \\ [\affilskip]
$^3$School of Physics and Astronomy, University of Birmingham, \\ Edgbaston, Birmingham B15 2TT, UK \\[\affilskip]
}

\pubyear{2012}
\volume{291}  %% insert here IAU Symposium No.
\jname{\mbox{Neutron Stars and Pulsars: Challenges and Opportunities after 80 years}}
\editors{J. van Leeuwen, ed.} 
\begin{document}

\maketitle

%% -- Abstract ----------------------------------
\begin{abstract}
The luminosity distribution of Galactic radio pulsars is believed to be
log-normal in form. Applying this functional form to populations of pulsars in
globular clusters,  we employ Bayesian methods to explore constraints on the
mean and standard deviation of the function, as well as the total number of
pulsars in the cluster. Our analysis is based on an observed number of pulsars
down to some limiting flux density, measurements of flux densities of
individual pulsars, as well as diffuse emission from the direction of the
cluster. We apply our analysis to Terzan 5 and demonstrate, under reasonable
assumptions, that the number of potentially observable pulsars is in a 95.45\%
credible interval of 133$^{+101}_{-58}$. Beaming considerations would increase
the true population size by approximately a factor of two.
\end{abstract}

% add below any authors, subjects and objects for indexing 
%   add more lines if necessary
%   but leave all lines commented out
%\index[author]{LastName1, Initials|textbf}
%\index[author]{LastName2, Initials|textbf}
%\index[subject]{Keyword1}
%\index[subject]{Keyword2}
%\index[object]{Object1}
%\index[object]{Object2}

\firstsection % if your document starts with a section,
\section{Introduction}

Globular clusters have high core stellar number densities that favour the
formation of low-mass X-ray binaries (LMXBs) that are believed to be the
progenitors of millisecond pulsars (MSPs; \cite[Alpar et al. 1982]{alp82}).
MSPs can be considered long-lived tracers of LMXBs, so constraints on the MSP
content provide unique insights into binary evolution and the integrated
dynamical history of globular clusters, while determining the radio luminosity
function of these pulsars helps shed light on their emission mechanism.

\cite[Faucher-Gigu\`ere \& Kaspi (2006)]{fau06} have shown that the luminosity
distribution of non-recycled Galactic pulsars appears to be log-normal in form.
More recently, \cite[Bagchi et al. (2011)]{bag11} have verified that the observed
luminosities of recycled pulsars in globular clusters are consistent with this
result. Assuming, therefore, that there is no significant difference between
the nature of Galactic and cluster populations, we use Bayesian techniques to
investigate some of the consequences that occur when one applies this
functional form to populations of pulsars in individual clusters. We
are interested in the situation where we observe $n$ pulsars with luminosities
above some limiting luminosity. There is a family of luminosity function
parameters ($\mu$, $\sigma$) and population sizes ($N$) that is consistent with
this observation, and here we analyze the posterior probabilities of different
members of this family given the data. In our case, the data are the
individual pulsar flux densities that we call $\{S_i\}$, the observed number of
pulsars, $n$ and the total diffuse flux density of the cluster, $S_{\rm obs}$.

\section{Bayesian parameter estimation} \label{sec_bayesian}

Luminosity and flux density are related by the standard pseudo-luminosity
equation $L = S~r^2$, where $r$ is the distance to the pulsar
(see \cite[Lorimer \& Kramer 2005]{lor05}). This implies that the luminosity
function is corrupted by uncertainties in distance. To mitigate this, we
decided to perform our analysis initially in terms of the measured flux
densities, and use a model of distance uncertainty to convert our results to
the luminosity domain. We take the distance to all pulsars in a cluster to be
the same. The log-normal in luminosity can then alternatively be written in
terms of flux density. The probability of detecting a pulsar with flux
density $S$ in the range $\log S$ to $\log S + d(\log S)$ is given by a
log-normal in $S$ as
\begin{equation}
p(\log S)~d(\log S) = \frac{1}{\sigma_S \sqrt{2 \pi}}e^{-\frac{(\log S - \mu_S)^2}{2 \sigma_S^2}}~d(\log S),
\end{equation}
where $S$ is in mJy, and $\mu_S$ and $\sigma_S$ are the mean and standard
deviation of the \emph{flux density distribution}.
The probability of observing a pulsar above the limit $S_{\rm min}$ is
then
\begin{equation}
\label{eq_p_obs}
p_{\rm obs} = \int_{\log S_{\rm min}}^{\infty} p(\log S)~d(\log S)
= \frac{1}{2}~{\rm erfc}\left(\frac{\log S_{\rm min} - \mu_S}{\sqrt{2} \sigma_S}\right).
\end{equation}

First, we consider as data the measured flux densities of pulsars in the
cluster, $\{S_i\}$. Ideally, the survey sensitivity limit
$S_{\rm min}$ can be taken as another datum, but its exact value is not always
known, so we decided to parametrize $S_{\rm min}$. The likelihood of observing
a set of pulsars with fluxes $\{S_i\}$ is represented as
\begin{equation}
\prod\limits_{i=1}^{n}p_i(\log S_i | \mu_S, \sigma_S, S_{\rm min})
 = \prod\limits_{i=1}^{n} \frac{1}{p_{\rm obs} \sigma_S \sqrt{2 \pi}}
e^{-\frac{(\log S_i - \mu_S)^2}{2 \sigma_S^2}}
\end{equation}
where $n$ is the number of observed pulsars in the cluster, and $p_{\rm obs}$
is as given in Equation~(\ref{eq_p_obs}). Uncertainties in the flux density
measurements are not considered here, but it has to be noted that it will have
the effect of underestimating the credible intervals on our posteriors.

To infer the total number of pulsars in the cluster, we follow
\cite[Boyles et al. (2011)]{boy11} to take as likelihood the probability of observing $n$ pulsars in
a cluster with $N$ pulsars, given by the binomial distribution
\begin{equation}
p(n | N, \mu_S, \sigma_S, S_{\rm min}) = \frac{N!}{n!(N-n)!}
~p_{\rm obs}^{n}~(1 - p_{\rm obs})^{N - n}.
\end{equation}

Next, we incorporate information about the observed diffuse flux
from the direction of the cluster. We assume that all radio emission
is due to the pulsars in the cluster, both resolved and unresolved. For the
likelihood of measuring the diffuse flux $S_{\rm obs}$, we choose
\begin{equation}
\label{eq_likelihood3}
p(S_{\rm obs} | N, \mu_S, \sigma_S) = \frac{1}{\sigma_{\rm diff} \sqrt{2 \pi}}
~e^{-\frac{(S_{\rm obs} - S_{\rm diff})^2}{2 \sigma_{\rm diff}^2}},
\end{equation}
where $S_{\rm diff}$ is the expectation of the total diffuse flux of a cluster
whose flux density distribution is a log-normal with parameters $\mu_S$ and $\sigma_S$,
and having $N$ pulsars, and $\sigma_{\rm diff}$ is the standard deviation.
Here, $S_{\rm diff} = N \langle S \rangle$ and
$\sigma_{\rm diff} = \sqrt{N}~{\rm SD}(S)$ where the expectation of $S$ is
given by $\langle S \rangle = 10^{\mu_S + \frac{1}{2} \sigma_S^2 \ln(10)}$
and the standard deviation of $S$,
${\rm SD}(S) = 10^{\mu_S + \frac{1}{2} \sigma_S^2 \ln(10)}
\sqrt{10^{\sigma_S^2 \ln(10)} - 1}$. We do not consider the uncertainty in the
diffuse flux measurement. The total likelihood,
$p(\log S_i, n, S_{\rm obs} | N, \mu_S, \sigma_S, S_{\rm min})$ is the product
of the three likelihoods computed above.

The flux density distribution of pulsars in a cluster is not suitable for
comparing the populations in different clusters, as it depends on the distance
to the cluster. So we transform the total likelihood obtained in the
previous subsection to the luminosity domain. Taking into account the
uncertainty in distance as a distribution of distances, $p(r)$, it can be shown
that the total likelihood in the luminosity domain is
\begin{equation}
p(\log S_i, n, S_{\rm obs} | N, \mu, \sigma, S_{\rm min}, r)
= p(\log S_i, n, S_{\rm obs} | N, \mu_S, \sigma_S, S_{\rm min}),
\end{equation}
where $\mu$ and $\mu_S$ are related additively by the term $2 \log r$, and
$\sigma$ and $\sigma_S$ are equal. The final joint posterior in luminosity is
then given by
\begin{equation}
\begin{split}
p(N, &\mu, \sigma, S_{\rm min}, r | \log S_i, n, S_{\rm obs})\\
\propto~&p(\log S_i, n, S_{\rm obs} | N, \mu, \sigma, S_{\rm min}, r)~p(N)~p(\mu)~p(\sigma)~p(S_{\rm min})~p(r).
\end{split}
\end{equation}
The prior on $N$ is taken to be uniform from $n$ to $\infty$.
We also use uniform priors on the model parameters $\mu$ and $\sigma$.
We choose a uniform prior on $S_{\rm min}$ in the range (0, min($S_i$)],
where the upper limit is the flux density of the least bright pulsar in the
cluster. The prior on $r$ is taken to be a Gaussian. This joint posterior is
integrated over various sets of model parameters to obtain marginalized posteriors.

\section{Applications}\label{sec_apps}

\begin{figure}[t]
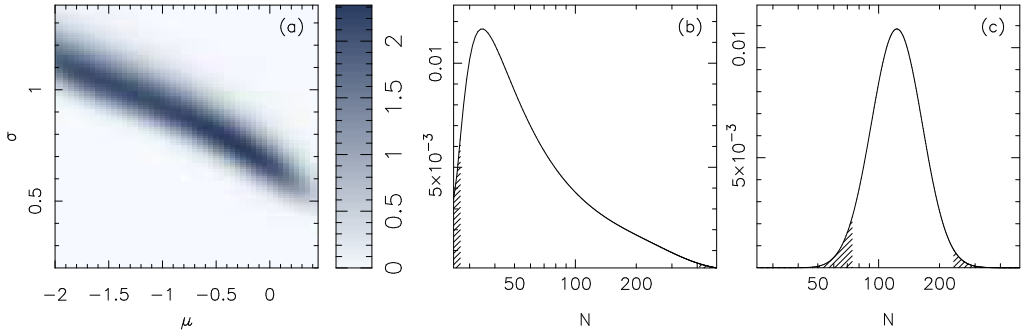

\begin{center}
\includegraphics[angle=-90,width=0.7\linewidth]{ter5_wide.ps}
\includegraphics[angle=-90,width=0.29\linewidth]{ter5_narrow.ps}
\caption{Results of the analysis for Terzan 5. For (a) and (b),
the analysis was run with wide priors on $\mu$ and $\sigma$, with the ranges
equal to those used by \cite[Bagchi et al.\ (2011)]{bag11}, their Figure 2.
(a) depicts the joint posterior on $\mu$ and $\sigma$, marginalized over $N$,
$S_{\rm min}$ and $r$; (b) is the marginalized posterior for $N$.
(c) Posterior on $N$ after applying narrow priors on $\mu$ and $\sigma$. In
(b) and (c), the shaded regions lie outside a 95.45\% credible interval.
\label{fig_results}}
\end{center}
\end{figure}

We applied our Bayesian technique to Terzan 5.
Although Terzan 5 has 34 known pulsars (Ransom S. M., private communication),
we take $n = 25$, the number of pulsars for which we have flux density measurements. The flux
densities of the individual pulsars were collected in a literature survey
(\cite[Bagchi et al. 2011]{bag11} and references therein).
The flux densities we used were scaled from those reported at 1950 MHz by
\cite[Ransom et al. (2005)]{ran05} and \cite[Hessels et al. (2006)]{hes06}
to 1400 MHz using a spectral index, $\alpha = -1.9$, using the power law
$S(\nu) \propto \nu^\alpha$. The observed diffuse flux density at 1400 MHz is
taken to be $S_{\rm obs} = 5.2$ mJy (\cite[Fruchter \& Goss 2000]{fru00}).
The prior on $N$ was chosen to be uniform in [$n$, 500], which is sufficiently
wide to ensure that the posterior does not rail against the prior boundaries.
We chose uniform distributions in the same range of $\mu$ and
$\sigma$ as used by \cite[Bagchi et al. (2011)]{bag11} as our priors.
We took $S_{\rm min}$ to be uniform in (0, min($S_i$)]. The most recent
measurement of the distance to Terzan 5, $r = 5.5\pm0.9$ kpc
(\cite[Ortolani et al. 2007]{ort07}), was used to model the distance prior as a
Gaussian. Figure~\ref{fig_results} shows the results of the analysis. The
median values of the three parameters with 95.45\% credible intervals are:
$N = 92^{+318}_{-64}$, $\mu = -0.9^{+1.2}_{-1.1}$ and $\sigma = 0.9^{+0.3}_{-0.4}$.

Note that $N$ is the size of the population of pulsars that are beaming towards
the Earth. Uncertainties notwithstanding, the beaming fraction of MSPs is generally
thought to be $>$ 50\% (\cite[Kramer et al.\ 1998]{kra98}). This, together with
the fact that most pulsars in globular clusters are MSPs, imply that the true
population size in a cluster is approximately a factor of two more than the
potentially observable population size.

\subsection{Using prior information}

In the framework developed in the previous section, we use broad uniform (non-informative)
priors for the mean and standard deviation of the log-normal.
This lack of prior information is apparent in Figure~\ref{fig_results}(b),
where $N$ is not very well constrained.
Prior information can help better constrain the parameters of interest.
\cite[Boyles et al.\ (2011)]{boy11} use models of Galactic pulsars from
\cite[Ridley \& Lorimer (2010)]{rid10} to narrow down $\mu$ to between
$-$1.19 and $-$1.04, and $\sigma$ to the range 0.91 to 0.98. We have chosen our
priors on $\mu$ and $\sigma$ to be uniform within these ranges. Applying the Bayesian
analysis over this narrower range of $\mu$ and $\sigma$ results
in much tighter constraints on $N$ as seen in Figure~\ref{fig_results}(c),
where $N = 133^{+101}_{-58}$. This result is consistent with that of
\cite[Bagchi et al.\ (2011)]{bag11}.

\section{Conclusions} \label{sec_conclusions}

The technique described here would be useful in future studies of the globular
cluster luminosity function where ongoing and future pulsar surveys are
expected to provide a substantial increase in the number of known pulsars
in many clusters. We anticipate that the increased amount of data would enable
us to constrain the distributions of $\mu$ and $\sigma$ independently (i.e.
without the need to assume prior information from the Galactic pulsar
population). Further interferometric measurements of the diffuse radio flux in
many clusters could provide improved constraints on $\mu$ and $\sigma$ by
measuring the flux contribution from the individually unresolvable population
of pulsars.


\begin{thebibliography}{}
\bibitem[Alpar \etal\ (1982)]{alp82}{Alpar, M. A.,
Cheng, A. F., Ruderman, M. A., \& Shaham, J.} 1982, \textit{Nature}, 300, 728
\bibitem[Bagchi \etal\ (2011)]{bag11}{Bagchi, M., Lorimer, D. R., \& Chennamangalam, J.}
2011, \textit{MNRAS}, 418, 477
\bibitem[Boyles \etal\ (2011)]{boy11}
{Boyles, J., Lorimer, D. R., Turk, P. J., Mnatsakanov, R., Lynch, R. S.,
Ransom, S. M., Freire, P. C., \& Belczynski, K.} 2011, \textit{ApJ}, 742, 51
\bibitem[Faucher-Gigu\`ere \& Kaspi (2006)]{fau06}
{Faucher-Gigu\`ere, C.-A., \& Kaspi, V. M.} 2006, \textit{ApJ}, 643, 332
\bibitem[Fruchter \& Goss (2000)]{fru00}
{Fruchter, A. S., \& Goss, W. M.} 2000, \textit{ApJ}, 536, 865
\bibitem[Hessels \etal\ (2006)]{hes06}
{Hessels, J. W. T., Ransom, S. M., Stairs, I. H., Freire, P. C. C., Kaspi, V. M.,
\& Camilo, F.} 2006, \textit{Science}, 311, 1901
\bibitem[Kramer \etal\ (1998)]{kra98}{Kramer, M.,
Xilouris, K. M., Lorimer, D. R., Doroshenko, O., Jessner, A., Wielebinski, R., 
Wolszczan, A., \& Camilo, F.} 1998, \textit{ApJ}, 501, 270
\bibitem[Lorimer \& Kramer (2005)]{lor05}{Lorimer, D. R.,
\& Kramer, M.} 2005, \textit{Handbook of Pulsar Astronomy}, Cambridge Univ. Press,
Cambridge, UK
\bibitem[Ortolani \etal\ (2007)]{ort07}
{Ortolani, S., Barbuy, B., Bica, E., Zoccali, M., \& Renzini, A.} 2007, \textit{A\&A}, 470, 1043
\bibitem[Ransom \etal\ (2005)]{ran05}{Ransom, S. M.,
Hessels, J. W. T., Stairs, I. H., Freire, P. C. C., Camilo, F., Kaspi, V. M.,
\& Kaplan, D. L.} 2005, \textit{Science}, 307, 892
\bibitem[Ridley \& Lorimer (2010)]{rid10}{Ridley, J. P.,
\& Lorimer, D. R.} 2010, \textit{MNRAS}, 404, 1081
\end{thebibliography}
\end{document}